\documentclass[runningheads]{llncs}
\usepackage{graphicx}
\usepackage{textcomp}
\usepackage{comment}
\usepackage{todonotes}
\usepackage{color}

\begin{document}
\title{Interpretable Deep Models for Cardiac Resynchronisation Therapy Response Prediction}     

\author{Esther Puyol-Ant\'on* \inst{1} \and 
Chen Chen  \inst{3} \and
James R. Clough \inst{1} \and 
Bram Ruijsink \inst{1,2} \and
Baldeep S. Sidhu \inst{1,2} \and
Justin Gould \inst{1,2} \and
Bradley Porter \inst{1,2} \and
Mark Elliott \inst{1,2} \and
Vishal Mehta \inst{1,2} \and
Daniel Rueckert \inst{3} \and
Christopher A. Rinaldi  \inst{1,2} \and
Andrew P. King \inst{1}}
\authorrunning{E Puyol-Ant\'on et al.}   
\institute{School of Biomedical Engineering \& Imaging Sciences, King\textquotesingle s College London, UK. \and  Guy’s and St Thomas\textquotesingle{} Hospital, London, UK. \and BioMedIA Group, Department of Computing, Imperial College London, UK}


\maketitle              

\begin{abstract} 
Advances in deep learning (DL) have resulted in impressive accuracy in some medical image classification tasks, but often deep models lack interpretability. The ability of these models to explain their decisions is important for fostering clinical trust and facilitating clinical translation. Furthermore, for many problems in medicine there is a wealth of existing clinical knowledge to draw upon, which may be useful in generating explanations, but it is not obvious how this knowledge can be encoded into DL models - most models are learnt either from scratch or using transfer learning from a different domain. In this paper we address both of these issues. We propose a novel DL framework for image-based classification based on a variational autoencoder (VAE). The framework allows prediction of the output of interest from the latent space of the autoencoder, as well as visualisation (in the image domain) of the effects of crossing the decision boundary, thus enhancing the interpretability of the classifier. Our key contribution is that the VAE disentangles the latent space based on `explanations' drawn from existing clinical knowledge. The framework can predict outputs as well as explanations for these outputs, and also raises the possibility of discovering new biomarkers that are separate (or disentangled) from the existing knowledge. We demonstrate our framework on the problem of predicting response of patients with cardiomyopathy to cardiac resynchronization therapy (CRT) from cine cardiac magnetic resonance images. The sensitivity and specificity of the proposed model on the task of CRT response prediction are 88.43\% and 84.39\% respectively, and we showcase the potential of our model in enhancing understanding of the factors contributing to CRT response.
\keywords{Interpretable ML  \and Cardiac Resynchronization Therapy \and Cardiac MRI \and Variational Autoencoder.}
\end{abstract}

\section{Introduction}
Deep learning (DL) methods have achieved near-human accuracy levels in some classification tasks in the medical domain. However, most DL methods operate as `black boxes', mapping a given input to a classification output, but offer little to no explanation as to how the output was decided upon. This has led to increasing research focus in recent years on techniques for interpretable DL \cite{doshi2017towards}. 
In healthcare, interpretations (or explanations) are important in promoting clinicians' and patients' trust in automated decisions, as well as potentially for legal reasons \cite{goodman2017european}.\\
Most existing work on interpretable DL has focused on ``post-hoc'' analysis of existing trained models \cite{guidotti2018survey}. Such approaches can be useful for visualising low-level features (e.g. parts of an image) that were important in producing the predicted output, but they do not offer a way to link these features with higher-level concepts that are well understood by clinicians. Furthermore, often there is a wealth of existing clinical knowledge that could be exploited in training a DL model to explain its decisions, but post-hoc analysis does not offer an easy way of linking this knowledge to the interpretation. An alternative, but less explored approach, which has the potential to address these limitations, is to include the need for an explanation into the training of the DL model. 
For example, Hind \textit{et al.,} \cite{hind2019ted} have proposed a generic framework for supervised machine learning that augments training data to include explanations elicited from domain experts. Similarly, Alvarez-Melis \textit{et al.,} \cite{melis2018towards} proposed a `self-explaining neural network' in which interpretability is built in architecturally and more interpretable concepts are learnt during training. Both of these methods offer advantages in terms of explanations, but they do not offer an obvious way to link low level features with higher level explanatory concepts. Our work is based upon this type of approach but we seek to extend it to enable links to be made between low-level features and higher level concepts. We believe that such an interpretable approach is an essential characteristic to promote clinician trust in DL models. We propose a framework based on a variational autoencoder (VAE) to map images to low-dimensional latent vectors. The primary classification task is performed using these latent vectors. In addition, we incorporate clinical domain knowledge by using secondary classifier(s) in the latent space to encourage disentanglement of explanatory concepts within the learnt representation. By using a VAE we are able to decode and visualise these concepts to provide an explanatory basis for the primary model output. Simultaneously, we are able to link the high level concepts to features in image space. The disentanglement based on existing domain concepts also allows us to use the model to learn new undiscovered biomarkers. \\
\indent \textbf{Related Work on Interpretable DL:} The idea of performing a classification task using fully connected layers from the latent space of a VAE has been proposed before. For example, Biffi \textit{et al.,} \cite{biffi2020explainable} adopted this approach to classify cardiac pathologies and used the decoder of the VAE to visualise the morphological features involved in the classification. Similarly, Clough \textit{et al.,} \cite{clough2019global} proposed a similar architecture to detect the presence of coronary artery disease and also used concept activation vectors to quantify the importance of different explanatory concepts. Our work is methodologically distinct from \cite{clough2019global,biffi2020explainable} as we incorporate secondary classifier(s) as a way of incorporating existing clinical knowledge into the model as well as providing explanations of outputs.\\

\indent \textbf{Cardiac Resynchronisation Therapy:}
We illustrate our approach on the important clinical task of predicting response to cardiac resynchronization therapy (CRT).
CRT is an established therapy for patients with medically refractory systolic heart failure and left ventricular dyssynchrony \cite{abraham2002cardiac}. Current consensus guidelines \cite{authors20132013} regarding selection for CRT focus on a limited set of patient characteristics including NYHA functional class, left ventricular ejection fraction (LVEF), QRS duration, type of bundle branch block, etiology of cardiomyopathy and atrial rhythm (sinus, atrial fibrillation).
However, using these guidelines approximately 30\% of patients do not respond to treatment \cite{mcalister2007cardiac},\cite{parsai2009toward}. The clinical research literature reveals a number of important insights into improving selection criteria. For example, it has been shown that strict left bundle branch block (LBBB) with type II contraction pattern is associated with increased response to CRT \cite{jackson2014u}. Another study demonstrated that the presence of septal flash (SF)\footnote{An inward-outward motion of the septum in early systole (mainly during isovolumetric contraction).} and apical rocking are also associated with improved response \cite{stankovic2016relationship}, \cite{marechaux2016role}.
A limited number of papers have investigated the use of machine learning to predict response to CRT. Peressutti \textit{et al.,} \cite{peressutti2017framework} used supervised multiple kernel learning (MKL) to combine motion information derived from cardiac magnetic resonance (CMR) imaging and non-motion data to predict CRT response, achieving approximately 90\% accuracy on a cohort of 34 patients. Cikes \textit{et al.,} \cite{cikes2019machine} used unsupervised MKL to combine echocardiographic data and clinical parameters to phenogroup patients with HF with respect to both outcomes and response to CRT. To the best of our knowledge, no DL models have been proposed for predicting response to CRT.

\textbf{Contributions:} We present a new DL image-based classification framework based on a VAE that enhances the interpretability of the classifier for the application of CRT response prediction. The main novelty lies in the use of secondary classifier(s) that enable links to be made between low level image features and higher level concepts, which we believe is an important prerequisite for clinical translation of DL based cardiac diagnosis tools.

\section{Materials}
\label{sec:materials}
A cohort of 73 patients fulfilling the conventional criteria for CRT was used in this study. The study was approved by the institutional ethics committee and all patients gave written informed consent. All patients underwent CMR imaging prior to CRT and 2D echocardiography imaging and clinical evaluation prior to CRT and at 6-month follow-up. CMR imaging was carried out on multiple scanners: Siemens Aera 1.5T, Siemens Biograph mMR 3T, Philips 1.5T Ingenia and Philips 1.5T and 3T Achieva. The CMR multi-slice short-axis (SA) stack was used in this study, which had a slice thickness between 8 and 10 mm, an in-plane resolution between 0.94x0.94mm\textsuperscript{2} to 1.5×1.5mm\textsuperscript{2} and a temporal resolution of $\sim$13-31 ms/frame. All patients were classified as responders or non-responders based on volumetric measures derived from 2D echocardiography acquired at the 6-month follow-up evaluation \cite{authors20132013}. Patients were classified as responders if they had a reduction of $\geq$15\% in left ventricular (LV) end-systolic volume after CRT, and non-responders otherwise. This information was used as the primary output label in training our proposed model. Based on a manual clinical assessment of the 2D echocardiography imaging acquired prior to CRT, those patients exhibiting SF were also identified by an experienced cardiologist. The SF information was used as an `explanatory concept' in training our proposed model to generate meaningful explanations. Note that this information would not normally be available at inference time, as it results from a time-consuming expert inspection. From the cohort, there were 47/73 patients who were responders to CRT. 27 of the 47 responders had SF and 10 of the 26 non-responders had SF. The ratio of SF responders/non-responders was 27/37 ($\sim$73\%), which is in line with the distribution reported by Parsai \textit{et al.,} \cite{parsai2009toward}
\section{Methods}
\label{sec:methods}
\begin{figure}[ht]%
    \centering
    \includegraphics[width=\textwidth]{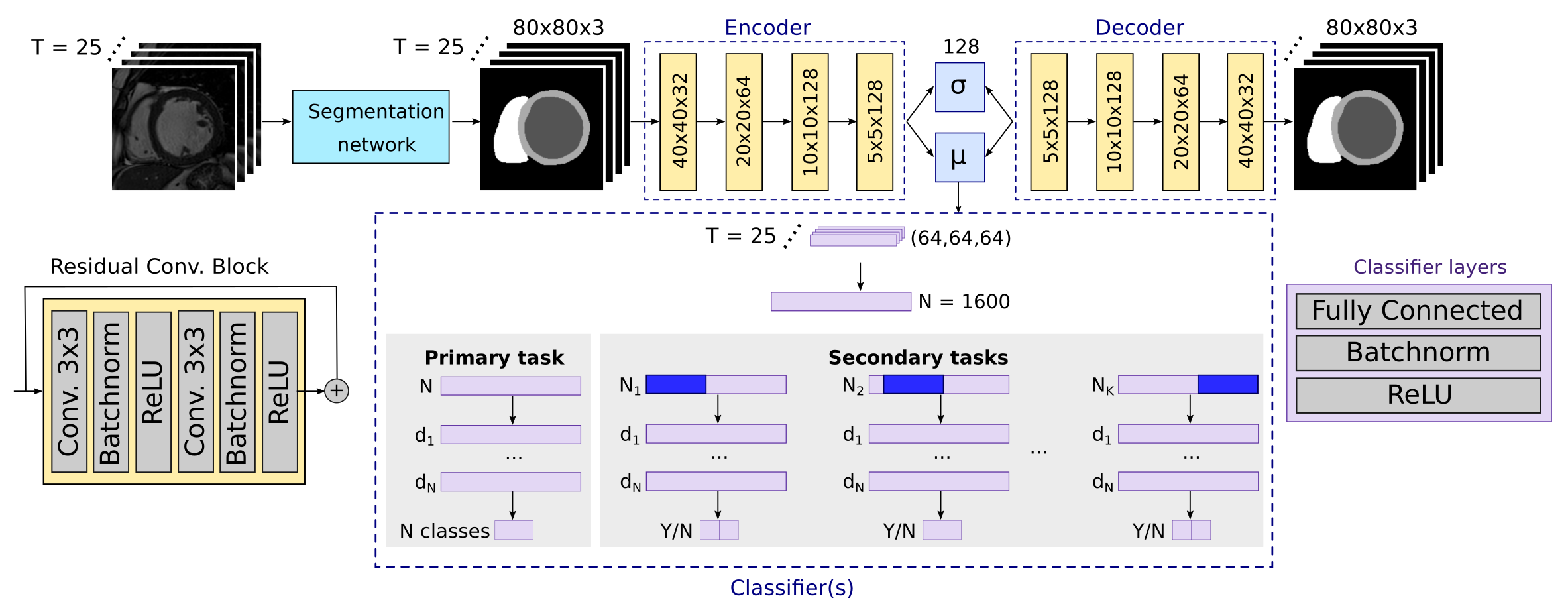}
    \caption{Diagram showing the architecture of the joint VAE/classification model. The inputs to the VAE are CMR segmentations from $T=25$ time points for the three top slices of the SA stack. The VAE consists of a series of residual convolutional blocks, with the image resolution and number of feature maps denoted in each block. For the multiple classification networks, the latent vectors belonging to the different time points are concatenated and used as inputs.}
    \label{fig:model}%
\end{figure}

Our CRT prediction model is illustrated in Figure \ref{fig:model}. The model consists of a segmentation network, followed by a VAE that incorporates multiple classifiers in the latent space. In the following sections we describe the different components of the pipeline.\\

\indent \textbf{Spatial-temporal normalisation} To correct for variation in acquisition protocols between vendors, all images were first resampled to an in-plane voxel size of 1.25 $\times$ 1.25$mm$, and temporally resampled to  $T=25$ frames per cardiac cycle using piecewise linear warping of cardiac timings.\\
\indent \textbf{Automatic segmentation network:} We used a U-net based architecture for automatic segmentation of the LV blood pool, LV myocardium and right ventricle (RV) blood pool from the SA slices in all frames through the cardiac cycle \cite{chen2019improving}. To take into account the inter-vendor differences in intensity distributions, the segmentation network was fine tuned with 300 images (multiple slices/time points from 20 CMR scans) from an independent clinical database.\\
\indent\textbf{VAE network:} The VAE model is composed of an encoder that compresses the data into a latent space (of dimensionality 128), followed by a decoder network trained to reconstruct the original data from the latent representations. The encoder/decoder architecture has three channels corresponding to CMR segmentations from the top three slices.\\
\indent\textbf{Primary task classifier:} Using the encoder network, the segmentations of the CMR images at all $T$ time points from each subject were mapped into $T$ latent vectors. The first fully connected layer of the classifier processes the latent vectors individually. The outputs of these layers are then concatenated into a single vector, which is used as input to three fully connected layers that predict CRT response for each subject.\\
\indent\textbf{Explanatory concept classifiers:} The clinical domain knowledge is incorporate as secondary classifier(s). In the experiments presented in this paper, we illustrate this idea using a single explanatory concept (SF). However, in principle multiple explanatory concepts could be used, each of which would use a different (possibly overlapping) portion of the latent space. We always ensure that a part of the latent space is unused by any of the secondary classifiers, to ensure that this represents `unknown' factors contributing to the primary output (i.e. CRT response). The secondary classifier(s) follow a similar structure to the primary task classifier.\\
We denote an input data sequence by $\mathbf{X}= [\mathbf{x}_1, \mathbf{x}_2, ... \mathbf{x}_T] $, and its corresponding latent mean and standard deviation vectors as $\mathbf{M}=[\mathbf{\mu}_1, \mathbf{\mu}_2, ... \mathbf{\mu}_T]$ and $\mathbf{\Sigma}=[\mathbf{\sigma}_1, \mathbf{\sigma}_2, ... \mathbf{\sigma}_T]$,
where $(\mathbf{\mu}_t, \mathbf{\sigma}_t) = \mathrm{Encoder}(\mathbf{x}_t)$. The decoded images are denoted by $\mathbf{\widetilde{X}}= [\mathbf{\tilde{x}}_1, \mathbf{\tilde{x}}_2, ... \mathbf{\tilde{x}}_T]$. For the primary classification task, the ground truth label is denoted by $y$ and the predicted label by $\tilde{y}=\mathrm{Classifier}(\mathbf{M})$, i.e. we use only the latent mean vector for classification. For the secondary tasks, we select a subset of the latent space $\mathbf{M_k}=[\mathbf{\mu}_{l_k}, \mathbf{\mu}_{l_k+1}, ... \mathbf{\mu}_{l_k+N_k}]$,
where $N_k<T$ is the size of the subset for secondary task $k$ and $l_k$ is the start of the subset for task $k$. This subset is used for classification of the explanatory concept(s) $\tilde{y_k}=\mathrm{Classifier}(\mathbf{M_k})$. The joint loss function for the VAE and the primary and secondary classifiers can then be written as follows:
\begin{equation}
    \mathcal{L}_{\mathrm{total}}
    = \frac{1}{T} \sum_{t=1}^{t=T} 
    \left[ \mathcal{L}_{\mathrm{re}}(\mathbf{x}_t, \mathbf{\tilde{x}}_t) 
    + \beta \mathcal{L}_{\mathrm{KL}}(\mathbf{\mu}_t, \mathbf{\sigma}_t) \right]
    + \gamma \mathcal{L}_{\mathrm{cl}}(y, \tilde{y}) 
    + \sum_{k=0}^{K}  \alpha_k \mathcal{L}_{\mathrm{cl}}(y_k, \tilde{y_k})
\end{equation}
where $\mathcal{L}_{\mathrm{re}}$ is the cross-entropy between the input segmentations and the output predictions,   $\mathcal{L}_{\mathrm{cl}}$ is the binary cross entropy loss for the classification tasks (primary/secondary), $\mathcal{L}_{\mathrm{KL}}$ is the Kullback-Leibler divergence between the latent variables and a unit Gaussian, and $\beta$, $\gamma$ and $\alpha_k$ are constants that weight the components of the loss function. $K$ is the number of secondary tasks and $y_k$ are their corresponding ground truth labels, which are provided as clinical domain knowledge (see Section~\ref{sec:materials}).

\section{Experiments and Results}
\label{sec:results} 
We applied the proposed pipeline for the primary classification task of predicting CRT response from pre-treatment CMR images. The explanatory concept used in the secondary classifier was SF. The inputs to the VAE were $80 \times 80$ segmentations of three basal slices of the SA stack, where each slice was treated as a channel of the network input. The data for each subject consist of $T=25$ segmentations per slice, representing one full cardiac cycle.\\
\indent  \textbf{Model Training:} The model was trained in three stages. First, we trained only using the VAE loss (i.e. $\gamma=0$ and $\alpha_k$ = 0) using 10,000 subjects (including both healthy and cardiovascular diseases) from the UK Biobank database \cite{petersen2015uk} for 500 epochs, and then fine tuned using CRT patient data for 300 epochs. Second, we trained both the VAE and the primary task classifier ($\alpha_k$=0) for 500 epochs. Finally, we trained the VAE, the main and secondary classifiers together for 300 epochs. We used a grid search strategy to identify the optimal beta that was selected as a trade-off between disentanglement and reconstruction quality. For the secondary classifier, we used the first half of the latent space ($N_0=64$ and $l_0=0$). Data augmentation was used during all phases of the training (random rotation and translation). The model was trained on a NVIDIA GeForce GTX TITAN X using Adam optimiser with learning rate equal to $\mathrm{10^{-4}}$ and batch size of 8. 
We set $\beta=0.2$, $\gamma=1$ and $\alpha_k=0.9$.  

\begin{table}[ht] 
\caption{Comparison between baseline, primary task VAE classifier and primary/secondary task VAE classifier. McNemar's test was used to compute the p-values.}
\centering
\begin{tabular}{lcccccccc}
\cline{2-9}
& \multicolumn{5}{c}{\textbf{CRT}} & \textbf{SF}\\ \cline{2-9}
Methods & BACC & SEN  & SPE & Dice & p-value &  BACC & SEN  & SPE   \\ \hline 
Baseline & 92.46 & 96.69 & 88.23 & - & - & - & - & - \\ 
VAE + CRT & 90.40 & 94.59 & 86.21 & 87.04 & 0.18 & - & - & -\\ 
VAE + CRT + SF & 86.41 & 88.43 & 84.39 & 85.85 & 0.06 & 80.24 & 73.42 & 87.06 \\ \hline 
\end{tabular}
\label{table:1}
\end{table}
\textbf{Classification and Reconstruction Results:} 
A 5-fold stratified cross-validation over the CRT patient data was employed to evaluate the proposed framework. For each fold, the decoder network was evaluated using the average Dice Score between the input segmentation and the predicted segmentation. The primary and secondary classifiers were evaluated using a receiver operating characteristic curve (ROC) analysis, and based on this the balanced accuracy (BACC), sensitivity (SEN) and specificity (SPE) were computed for the optimal classifier selected using the Youden index.
For comparison, we also evaluated a `baseline' classifier (i.e. encoder + CRT classifier only) and also a version of our model featuring only the primary task classifier and no secondary task (i.e. VAE + primary classifier).
Table \ref{table:1} summarises the results for the proposed method and the comparative methods. To compare the performance of the different classification algorithms we used McNemar's test between the baseline method and the other methods. \\

\begin{figure}[h]%
  \includegraphics[width=\textwidth]{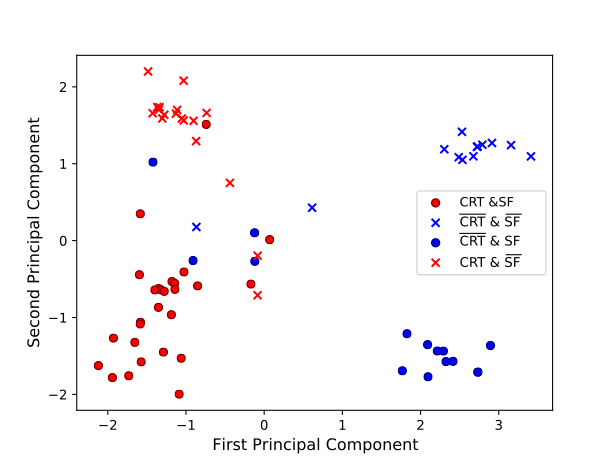}
    \caption{PCA of the latent space vectors for the test cases, where red are CRT responders and blue non CRT responders, dots are subjects with SF and crosses subjects without SF}%
    \label{fig:latent_space}%
\end{figure}

\textbf{Disentanglement of the Latent Space:} 
Figure \ref{fig:latent_space} shows the first two PCA components of the latent space on the test database. It is visible that the primary and secondary classifier has enforced the disentanglement of the latent space between CRT responders and non responders. 

\begin{figure}[ht]
    \centering
    \includegraphics[width=\textwidth]{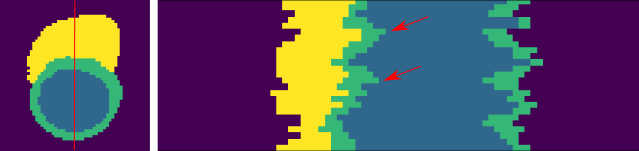}
    \caption{Two dimensional M-Mode visualisation of the mid LV and RV over the full cardiac cycle. Red line indicates profile line selected to generate the two dimensional M-Mode image. Arrows indicate irregular motion of the septum during contraction, which likely corresponds to septal flash.}
    \label{fig:sf}
\end{figure}

\textbf{Visualisation of the Explanatory Concepts:} One of the strengths of our proposed model is that it enables visualisation in the image domain of the secondary classification task to investigate if the learned features correspond to the clinical domain knowledge. To illustrate this concept, we computed the mean point in the latent space from all subjects classified as having SF and reconstructed an image sequence using the VAE decoder. Figure \ref{fig:sf} shows a two-dimensional M-Mode representation of a line crossing the LV and RV through the septum of the reconstructed images. The video of the full cardiac cycle is available in the supplementary material. An experienced cardiologist reviewed the video and the M-Mode representation and reported that it shows an irregular motion of the septal wall during contraction, most likely indicating SF.

\section{Discussion}
\label{sec:discussion}
In this work we propose a model that not only performs classification, but also allows interpretation of features important in the classification. We achieve state-of-the-art performance for CRT response prediction, and this is the first time that DL has been used for this purpose. Our DL model is fully automated and requires no user interaction to generate the model input, unlike previous machine learning approaches for CRT response prediction \cite{cikes2019machine,peressutti2017framework}. Our key novelty is that we use additional secondary classifier(s) to encode existing clinical knowledge into the model to enable it to explain its output. Our results show that our explanatory classifier has similar performance to the baseline VAE, but offers the possibility to disentangle the latent space based on clinical knowledge and hence explain its decisions to clinicians.

To illustrate the ability of our method to incorporate clinical knowledge we showed in Figure \ref{fig:sf} that our model had learnt in a weakly supervised manner the concept of SF, which is associated with positive response to CRT treatment. In future work we aim to investigate extending the current framework to incorporate multiple explanatory concepts. Apart from allowing interpretation and thus improving trust, encoding of clinical concepts into the model offers a way to incorporate the extensive biophysical knowledge that is already known about a disease. It is likely that this will in the long run improve DL applicability. Furthermore, the proposed model has the potential to discover new explanatory factors related to CRT response by using the VAE decoder to visualise other portions of the latent space. 

We used segmentations instead of intensity images to train our model as we achieved  slightly higher performance in this way and the quality of reconstructed images and latent space interpolations was superior. However, segmentations do not offer the same level of detail as CMR intensity images and therefore this could cause a loss of information that could be important in predicting CRT response. In the future we will investigate the use of an adversarial loss to ensure high-quality intensity image reconstructions which can then be used to visualise both structural and textural features relevant to the classification.

Finally, the proposed model currently concatenates the time series of latent vectors. A further possible extension would be to use more sophisticated architectures such as recurrent neural networks that take advantage of the temporal correlations between frames.

\section*{Acknowledgements}
This work was supported by the EPSRC (EP/R005516/1 and EP/P001009/1) and the Wellcome EPSRC Centre for Medical Engineering at the School of Biomedical Engineering and Imaging Sciences, King’s College London (WT 203148/Z/16/Z). This research has been conducted using the UK Biobank Resource under Application Number 17806.

\bibliographystyle{splncs04}
\bibliography{refs}

\begin{thebibliography}{10}
\providecommand{\url}[1]{\texttt{#1}}
\providecommand{\urlprefix}{URL }
\providecommand{\doi}[1]{https://doi.org/#1}

\bibitem{abraham2002cardiac}
Abraham, W.T., Fisher, W.G., Smith, A.L., Delurgio, D.B., Leon, A.R., Loh, E.,
  Kocovic, D.Z., Packer, M., Clavell, A.L., Hayes, D.L., et~al.: Cardiac
  resynchronization in chronic heart failure. New England Journal of Medicine
  \textbf{346}(24),  1845--1853 (2002)

\bibitem{biffi2020explainable}
Biffi, C., Cerrolaza, J.J., Tarroni, G., Bai, W., De~Marvao, A., Oktay, O.,
  Ledig, C., Le~Folgoc, L., Kamnitsas, K., Doumou, G., et~al.: Explainable
  anatomical shape analysis through deep hierarchical generative models. IEEE
  Transactions on Medical Imaging  (2020)

\bibitem{chen2019improving}
Chen, C., Bai, W., Davies, R.H., Bhuva, A.N., Manisty, C., Moon, J.C., Aung,
  N., Lee, A.M., Sanghvi, M.M., Fung, K., et~al.: Improving the
  generalizability of convolutional neural network-based segmentation on cmr
  images. arXiv preprint arXiv:1907.01268  (2019)

\bibitem{cikes2019machine}
Cikes, M., Sanchez-Martinez, S., Claggett, B., Duchateau, N., Piella, G.,
  Butakoff, C., Pouleur, A.C., Knappe, D., Biering-S{\o}rensen, T., Kutyifa,
  V., et~al.: Machine learning-based phenogrouping in heart failure to identify
  responders to cardiac resynchronization therapy. European journal of heart
  failure  \textbf{21}(1),  74--85 (2019)

\bibitem{clough2019global}
Clough, J.R., Oksuz, I., Puyol-Ant{\'o}n, E., Ruijsink, B., King, A.P.,
  Schnabel, J.A.: Global and local interpretability for cardiac mri
  classification. In: International Conference on Medical Image Computing and
  Computer-Assisted Intervention. pp. 656--664. Springer (2019)

\bibitem{doshi2017towards}
Doshi-Velez, F., Kim, B.: Towards a rigorous science of interpretable machine
  learning. arXiv preprint arXiv:1702.08608  (2017)

\bibitem{goodman2017european}
Goodman, B., Flaxman, S.: European union regulations on algorithmic
  decision-making and a “right to explanation”. AI magazine
  \textbf{38}(3),  50--57 (2017)

\bibitem{guidotti2018survey}
Guidotti, R., Monreale, A., Ruggieri, S., Turini, F., Giannotti, F., Pedreschi,
  D.: A survey of methods for explaining black box models. ACM computing
  surveys (CSUR)  \textbf{51}(5),  1--42 (2018)

\bibitem{hind2019ted}
Hind, M., Wei, D., Campbell, M., Codella, N.C., Dhurandhar, A., Mojsilovi{\'c},
  A., Natesan~Ramamurthy, K., Varshney, K.R.: Ted: Teaching ai to explain its
  decisions. In: Proceedings of the 2019 AAAI/ACM Conference on AI, Ethics, and
  Society. pp. 123--129 (2019)

\bibitem{jackson2014u}
Jackson, T., Sohal, M., Chen, Z., Child, N., Sammut, E., Behar, J., Claridge,
  S., Carr-White, G., Razavi, R., Rinaldi, C.A.: A u-shaped type ii contraction
  pattern in patients with strict left bundle branch block predicts
  super-response to cardiac resynchronization therapy. Heart Rhythm
  \textbf{11}(10),  1790--1797 (2014)

\bibitem{marechaux2016role}
Marechaux, S., Menet, A., Guyomar, Y., Ennezat, P.V., Guerbaai, R.A., Graux,
  P., Tribouilloy, C.: Role of echocardiography before cardiac
  resynchronization therapy: new advances and current developments.
  Echocardiography  \textbf{33}(11),  1745--1752 (2016)

\bibitem{mcalister2007cardiac}
McAlister, F.A., Ezekowitz, J., Hooton, N., Vandermeer, B., Spooner, C.,
  Dryden, D.M., Page, R.L., Hlatky, M.A., Rowe, B.H.: Cardiac resynchronization
  therapy for patients with left ventricular systolic dysfunction: a systematic
  review. Jama  \textbf{297}(22),  2502--2514 (2007)

\bibitem{melis2018towards}
Melis, D.A., Jaakkola, T.: Towards robust interpretability with self-explaining
  neural networks. In: Advances in Neural Information Processing Systems. pp.
  7775--7784 (2018)

\bibitem{authors20132013}
Members, A.F., Brignole, M., Auricchio, A., Baron-Esquivias, G., Bordachar, P.,
  Boriani, G., Breithardt, O.A., Cleland, J., Deharo, J.C., Delgado, V.,
  et~al.: 2013 esc guidelines on cardiac pacing and cardiac resynchronization
  therapy: the task force on cardiac pacing and resynchronization therapy of
  the european society of cardiology (esc). developed in collaboration with the
  european heart rhythm association (ehra). European heart journal
  \textbf{34}(29),  2281--2329 (2013)

\bibitem{parsai2009toward}
Parsai, C., Bijnens, B., Sutherland, G.R., Baltabaeva, A., Claus, P.,
  Marciniak, M., Paul, V., Scheffer, M., Donal, E., Derumeaux, G., et~al.:
  Toward understanding response to cardiac resynchronization therapy: left
  ventricular dyssynchrony is only one of multiple mechanisms. European Heart
  Journal  \textbf{30}(8),  940--949 (2009)

\bibitem{peressutti2017framework}
Peressutti, D., Sinclair, M., Bai, W., Jackson, T., Ruijsink, J., Nordsletten,
  D., Asner, L., Hadjicharalambous, M., Rinaldi, C.A., Rueckert, D., et~al.: A
  framework for combining a motion atlas with non-motion information to learn
  clinically useful biomarkers: application to cardiac resynchronisation
  therapy response prediction. Medical image analysis  \textbf{35},  669--684
  (2017)

\bibitem{petersen2015uk}
Petersen, S.E., Matthews, P.M., Francis, J.M., Robson, M.D., Zemrak, F.,
  Boubertakh, R., Young, A.A., Hudson, S., Weale, P., Garratt, S., et~al.: Uk
  biobank’s cardiovascular magnetic resonance protocol. Journal of
  cardiovascular magnetic resonance  \textbf{18}(1), ~8 (2015)

\bibitem{stankovic2016relationship}
Stankovic, I., Prinz, C., Ciarka, A., Daraban, A.M., Kotrc, M., Aarones, M.,
  Szulik, M., Winter, S., Belmans, A., Neskovic, A.N., et~al.: Relationship of
  visually assessed apical rocking and septal flash to response and long-term
  survival following cardiac resynchronization therapy (predict-crt). European
  Heart Journal-Cardiovascular Imaging  \textbf{17}(3),  262--269 (2016)

\end{thebibliography}

\end{document}